\documentclass[twocolumn]{elsart3p}

\usepackage{graphicx}

\usepackage{amssymb}

\begin{document}
\bibliographystyle{h-elsevier}

\begin{frontmatter}



\title{Performance and Fundamental Processes at Low Energy in a Two-Phase Liquid Xenon Dark Matter Detector}


\author[case]{T. Shutt\corauthref{tom}}
\author[ptown]{C. E. Dahl}
\author[ptown]{J. Kwong}
\author[case]{A. Bolozdynya}
\author[case]{P. Brusov}
\corauth[tom]{tshutt@case.edu}
\address[case]{Case Western Reserve University, Physics Dept, Cleveland, OH  44106}
\address[ptown]{Princeton Univeristy, Physics Dept, Princeton, NJ  08540}

\begin{abstract}
We extend the study of the performance of a prototype two-phase liquid xenon WIMP dark matter detector to recoil energies below 20 keV.   We demonstrate a new method for obtaining the best estimate of the energies of events using a calibrated sum of charge and light signals and introduce the corresponding discrimination parameter, giving its mean value at 4~kV/cm for electron and nuclear recoils up to 300 and 100~keV, respectively.  We show that fluctuations in recombination limit discrimination for most energies, and reveal an improvement in discrimination below 20 keV due to a surprising increase in ionization yield for low energy electron recoils.  This improvement is crucial for a high-sensitivity dark matter search.
\end{abstract}

\end{frontmatter}

\section{Introduction and Detector Description}
\label{intro}

Two-phase time-projection-chambers are a powerful technique for combining the high mass of a liquid-phase detector with the highly-sensitive gas-phase readout of electrons via proportional scintillation light \cite{Bolozdynya:1999dualphase}.   Recently, there has been a large effort to develop Xe and Ar two-phase detectors to search for WIMP dark matter \cite{Smith:2003zeplin,Kudryavtsev:2003zeplin,Yamashita:2003xmassdm,Brunetti:2004warp,CaseBrownColumbia:2006NucRec}, including by the XENON collaboration \cite{Aprile:2002xenon}, in which our group is a member.  These detectors consist of a liquid target with a gas region above and grid structures to create electric fields which drift electrons to the liquid surface, extract them into the gas phase, and induce them to create proportional scintillation (``S2'').   The initial (``S1'') scintillation light from the event site and the S2 light are measured by PMTs; the time difference between these signals gives the event depth.  A key feature of these detectors is discrimination between electron recoils (from radioactive backgrounds) and nuclear recoils (from WIMPs) based on the different amounts of charge recombination for these two types of interactions.  This was recently demonstrated in prototype detectors by our group and by the Columbia and Brown XENON groups \cite{CaseBrownColumbia:2006NucRec}.   This note extends these measurements to below 5~keV, explores the dominant role played by fluctuations in recombination, and reveals a surprising improvement in discrimination below 20 keV.

Our detector has an active liquid region 3.7 cm $\O$ and 0.96 cm deep, with stretched wire grids to define the drift field and a 10 kV/cm field region in the gas phase.  Two 178~nm xenon scintillation-light sensitive PMTs  (Hamamatsu 9288) are located below and above the active volume.  We estimate a light collection efficiency of $\sim$50\%, resulting in $\sim$1~photoelectrons (pe) per keV for nuclear recoils ($\sim$5~pe/keV for electron recoils at zero field).   The trigger threshold was $\sim$5 electrons (in the S2 signal), with $\sim$50\% acceptance for S1 single photo electrons.  The detector employs a temperature-controlled, liquid nitrogen coldfinger-cooled cryostat, with $\leqslant \pm0.2$~K stability.  The level of the liquid surface was measured using three parallel plate capacitors, allowing alignment of the liquid level and grids to $\sim 0.1^{\circ}$.   The detector was calibrated with 122~keV gammas from $^{57}$Co, and discrimination measurements were made using gammas from \(^{133}\)Ba and neutrons from a 25 $\mu$Ci \(^{252}\)Cf fission source.   Inelastic neutron reactions gave a prompt 40~keV line from  \(^{129}\)Xe.  Purity of the Xe, maintained with a heated getter purifier, was measured to $\lesssim$3\% charge loss over 1~cm via the depth dependence of the charge signal for 122 keV gammas.

\section{Recombination and a Better Method of Determining Energy} 
\label{combined_scale}

The charge-to-light anti-correlation  in which each electron lost to recombination leads to an additional scintillation photon is of fundamental importance to the response of the detector, and has long been recognized \cite{Doke:2002scintyields,Conti:2003anticorrelation}.   In this paper, we assume this anti-correlation to strictly hold (which implies no bi-excitonic quenching as suggested for very dense alpha-tracks by Hitachi \cite{Hitachi:1992quenching}), and find a number of interesting consequences, not yet discussed in the literature.  The first is an absolute calibration of the number of of scintillation (S1) photons emitted ($n_\gamma$), given the absolute number of electrons measured with S2 ($n_e$), by setting $\Delta n_e = \Delta n_{\gamma}$ for a peak where the mean recombination is shifted by varying the drift field.  We followed this procedure using the ${}^{57}$Co 122~keV gamma at drift fields from 200~V/cm to 2~kV/cm.  Before this work (also reported in \cite{CaseBrownColumbia:2006NucRec}) there were no published data for the charge (or light) yield of electron recoils below 500~keV, so we first calibrated $n_e$ as a function of drift field by direct measurement in the liquid phase with a calibrated charge pre-amplifier.  The light yield calibration so obtained agrees to $\sim$10\% with a MonteCarlo calculation of the light collection efficiency.

\begin{figure}[!t]
\begin{center}
	\includegraphics[width =3.25in]{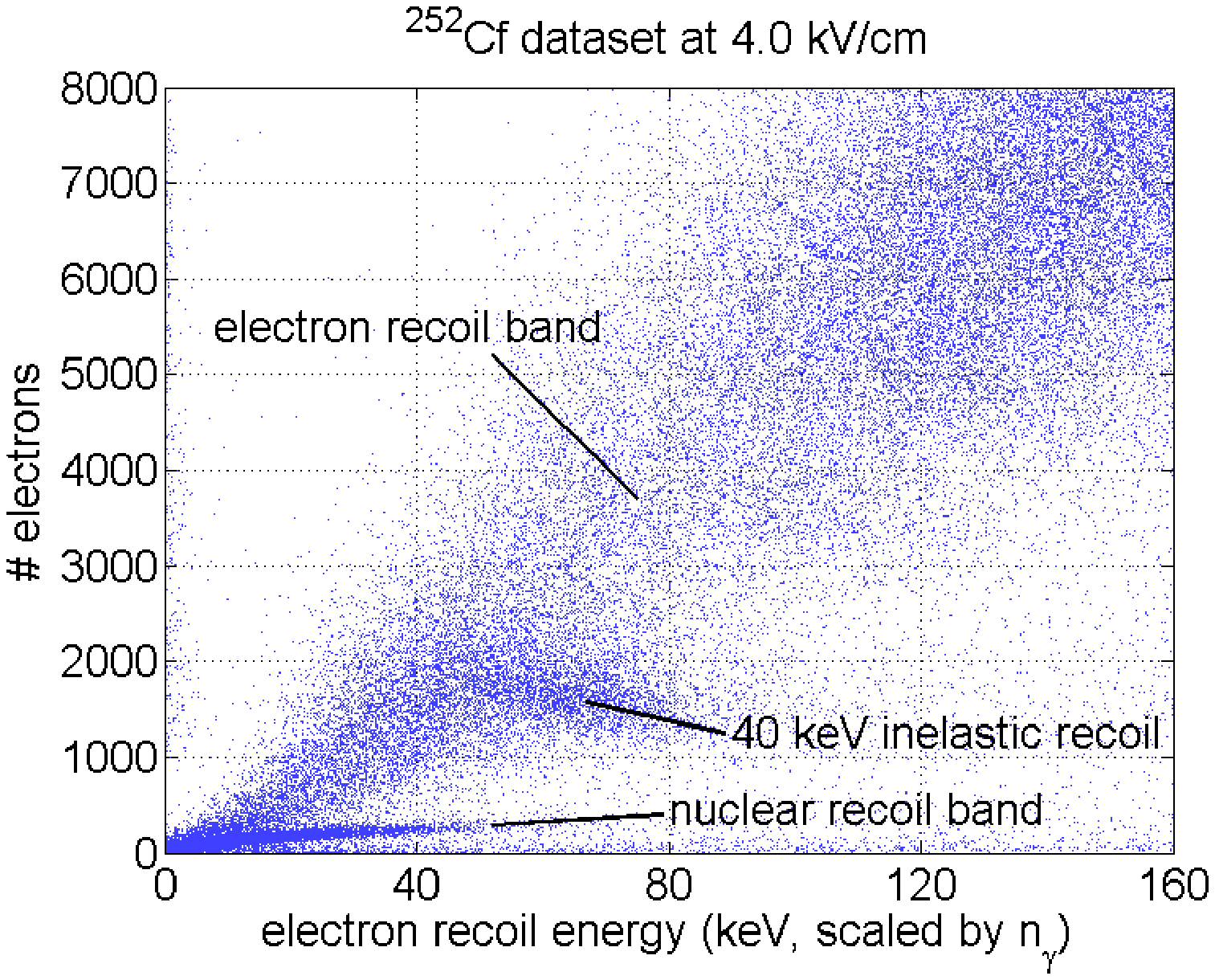} 
	\includegraphics[width =3.25in]{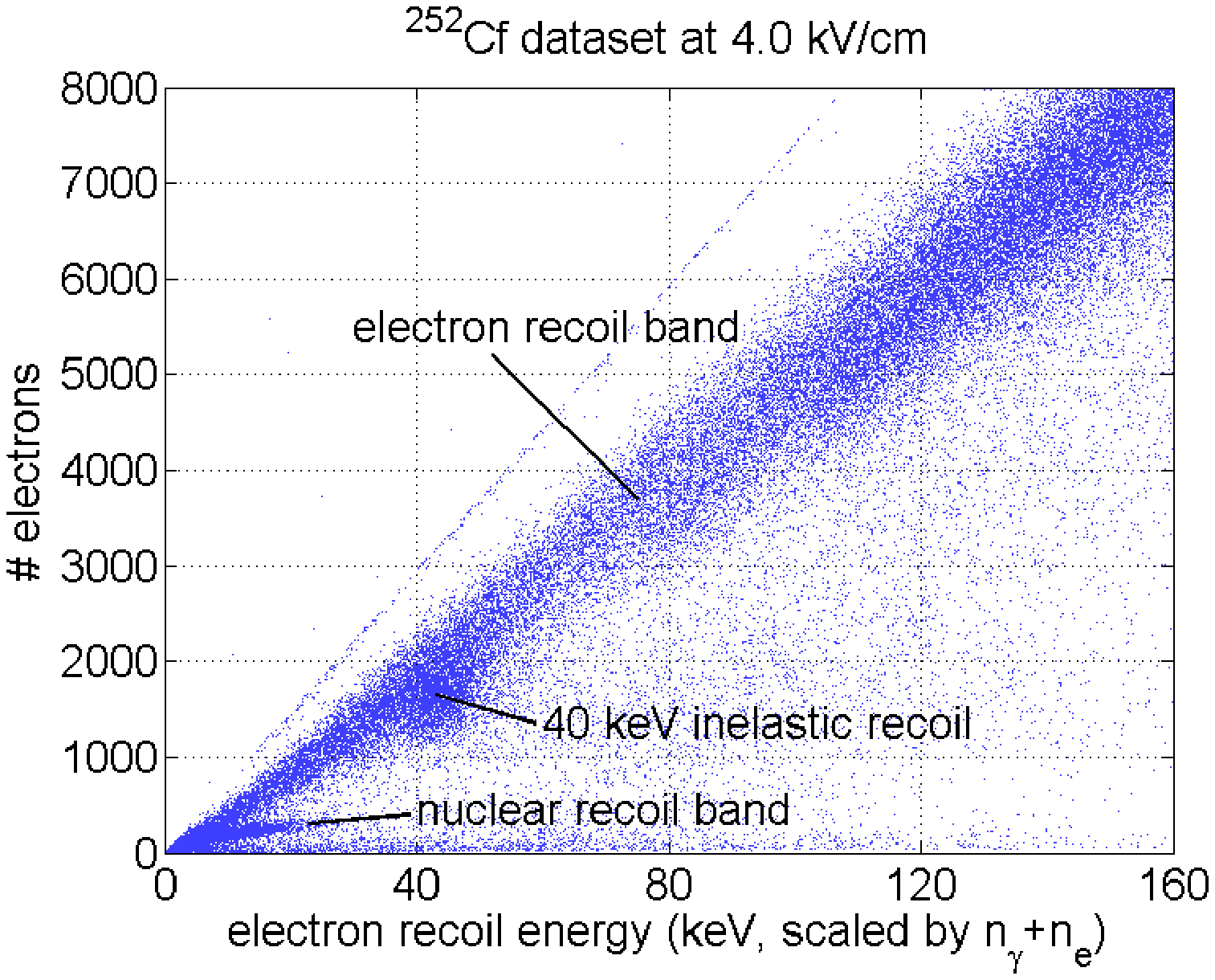} 
\end{center}
\caption{\label{fig:2D_resolution}The response of the detector to both gammas and neutrons from a $^{252}$Cf neutron source, showing the different bands for nuclear and electronic recoils.  The number of electrons, $n_e$, is plotted against energy (keVee) derived from \emph{above} scintillation alone, and \emph{below} the calibrated sum of scintillation and charge.}
\end{figure}

We then recognize $E=(n_\gamma+n_e)\cdot W$ as the best estimate of the energy of the event, where $W$ (the `W-value') is the average energy required to create any excitation  (photon or electron), found by the measurement above to be $13.46\pm0.29$~eV.  This new energy scale has several advantages over the standard calibration based on scintillation alone.   First, it requires no drift-field dependent calibration factors.    Moreover, it removes non-linearities in the scintillation calibration due to the strong energy dependence of recombination.  This is seen in Fig. \ref{fig:2D_resolution} in the position of the 40~keV peak in the two energy scales, both calibrated at 122~keV.   With this new scale, the energy scaling between nuclear recoil energy (keVr) and electron recoil energy (keVee) is given only by the  `Lindhard factor' \cite{Lindhard:1963nuclearsupression} describing the suppression of electronic  excitations by nuclear recoils relative to electron recoils (for Xe it is calculated to be a mildly energy-dependent factor of $\sim$4).   Finally, fluctuations in the amount of recombination, which have long been known to dominate the resolution of ionization measurements in liquid xenon, are completely removed from this energy scale, greatly improving the energy resolution as is evident in Fig. \ref{fig:2D_resolution}.  (Note, however, that this transformation does not affect discrimination).  These fluctuations are believed to stem from fluctuations in the electron recoil track structure which lead to different density distributions in the electron-ion charge clouds.  These are evidently the dominant dispersive mechanism in the electron recoil band.  From the resolutions in S1, S2, and the new combined scale (16\%, 10\% and 5.7\%  $\sigma /mean$, respectively at 122 keV), one can determine the magnitude of recombination fluctuations.  In particular, if the only correlated fluctuations in S1 and S2 are due to recombination, then $\sigma^2_{n_x,recomb}=\frac{1}{2}(\sigma^2_{n_\gamma}+\sigma^2_{n_e} - \sigma^2_{n_\gamma + n_e})$, where $\sigma^2_{n_x,recomb}$ is the variance in the ionization and scintillation signals due to recombination fluctuations.

\section{Low energy discrimination} \label{nuclear_recoils}

\begin{figure}[!t]
\begin{center}
	\includegraphics[width =3.25in]{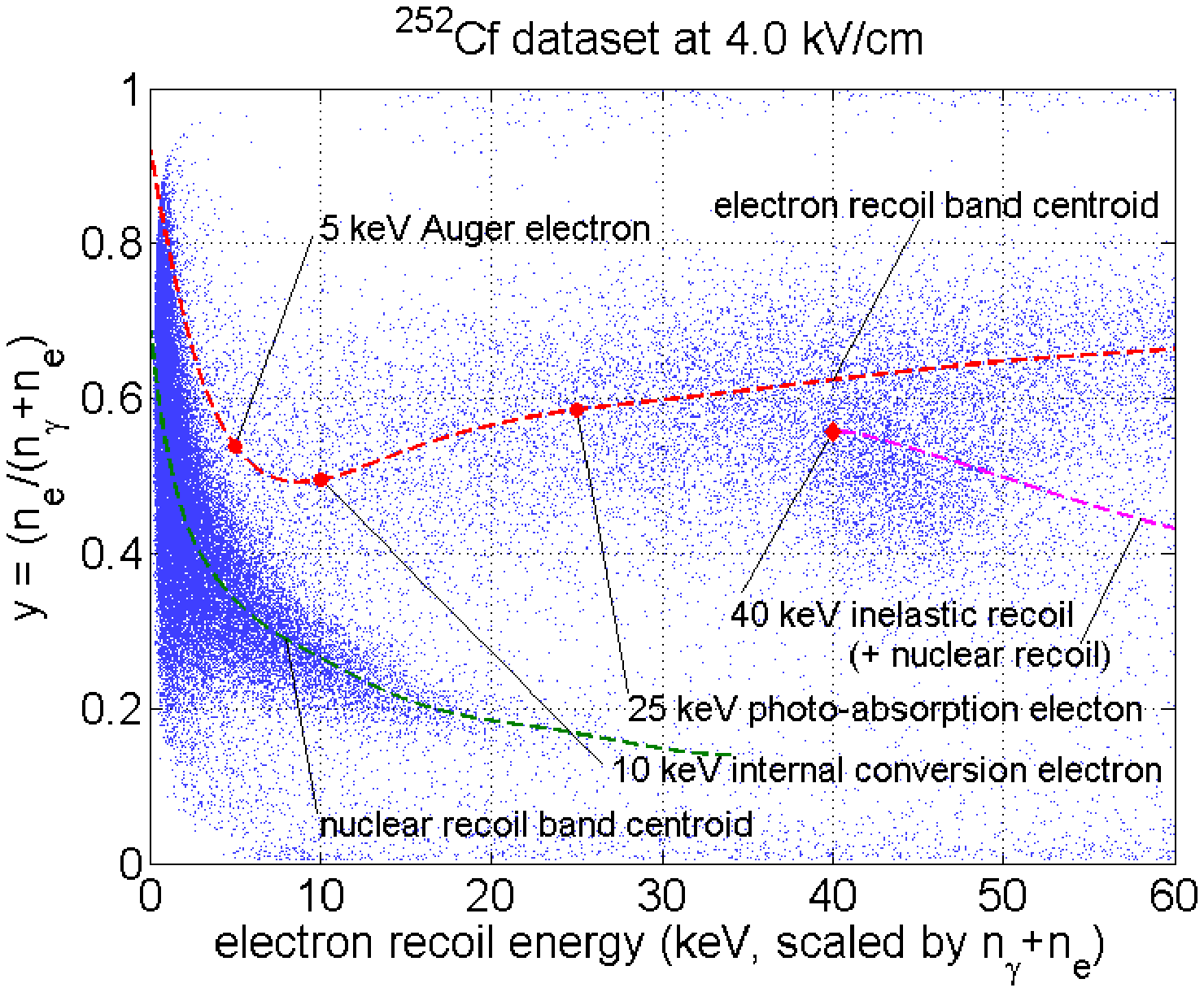}
	\includegraphics[width =3.25in]{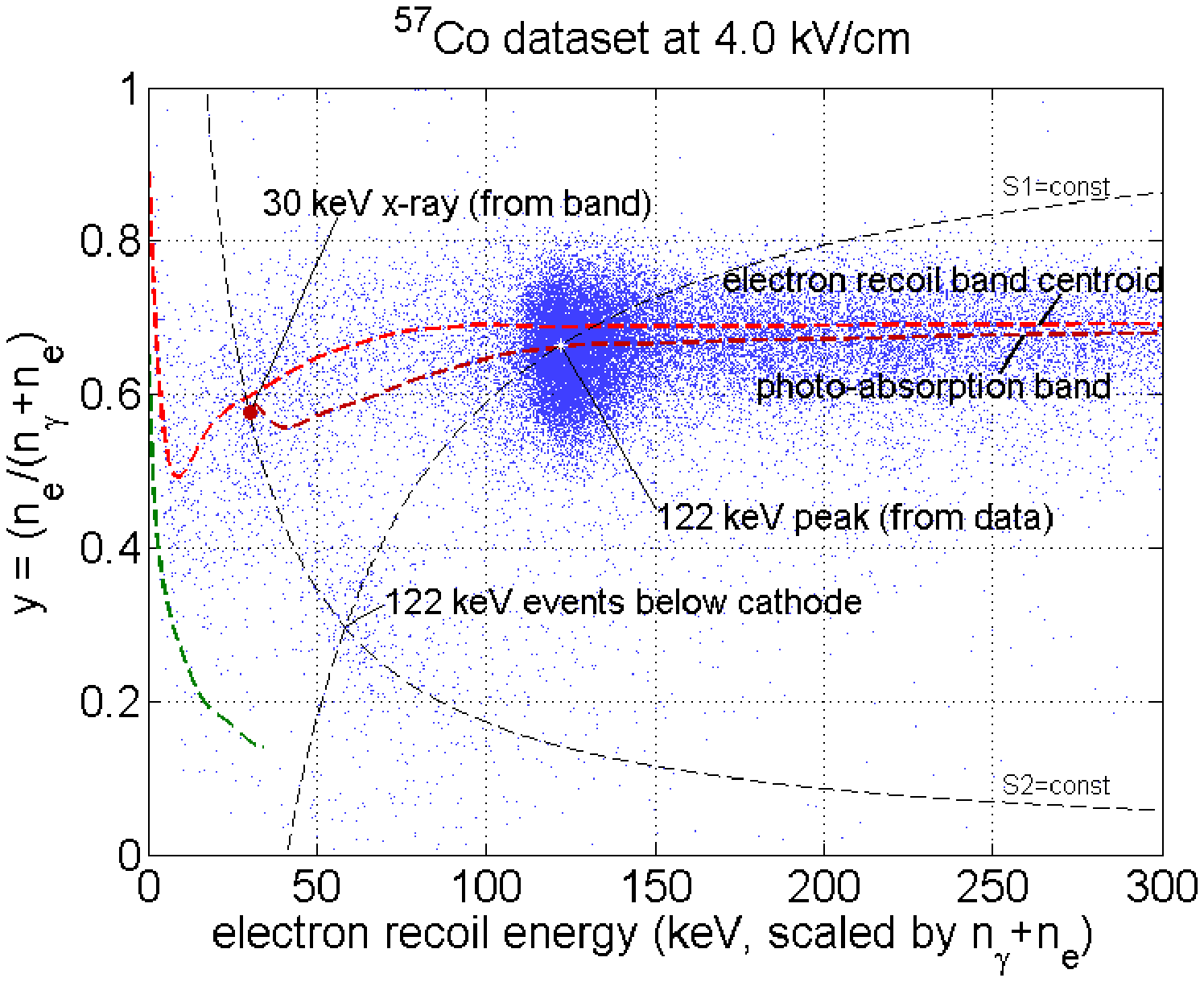}
\end{center}
\caption{\label{fig:Bands_w_40_keV}Electron and nuclear recoil band centroids laid over data taken with sources ${}^{252}$Cf \emph{above} and ${}^{57}$Co \emph{below}.  The bands for inelastic nuclear recoils and photo-absorption events are derived from the raw electron and nuclear recoil band centroids.  The inelastic band is given by a nuclear recoil plus a prompt 40~keV decay, which produces a 10~keV internal conversion electron from the K-shell plus a 30~keV x-ray, which in turn is photo-absorbed in the L-shell of a nearby atom, giving a 25~keV electron and 5~keV Auger electron.  The band for photo-absorption of gammas in the K-shell is similarly calculated.  As expected, the inelastic band goes through the observed inelastic recoil peak \emph{above}, and the 122~keV peak from ${}^{57}$Co also lies on the derived photo-absorption band \emph{below}.  Also drawn in below are lines of constant S1 and S2, going through the 122~keV peak and 30~keV x-ray position respectively.  These identify a population of events where the initial 122~keV photo-absorption occurred below the cathode, giving S1 but no S2, while the following 30~keV x-ray traveled above the cathode before being captured.  The drift times for these events (based on the 30~keV S2) confirm that they occur near the cathode.}

\end{figure}

\begin{figure}[!t]
\begin{center}
	\includegraphics[width =3.25in]{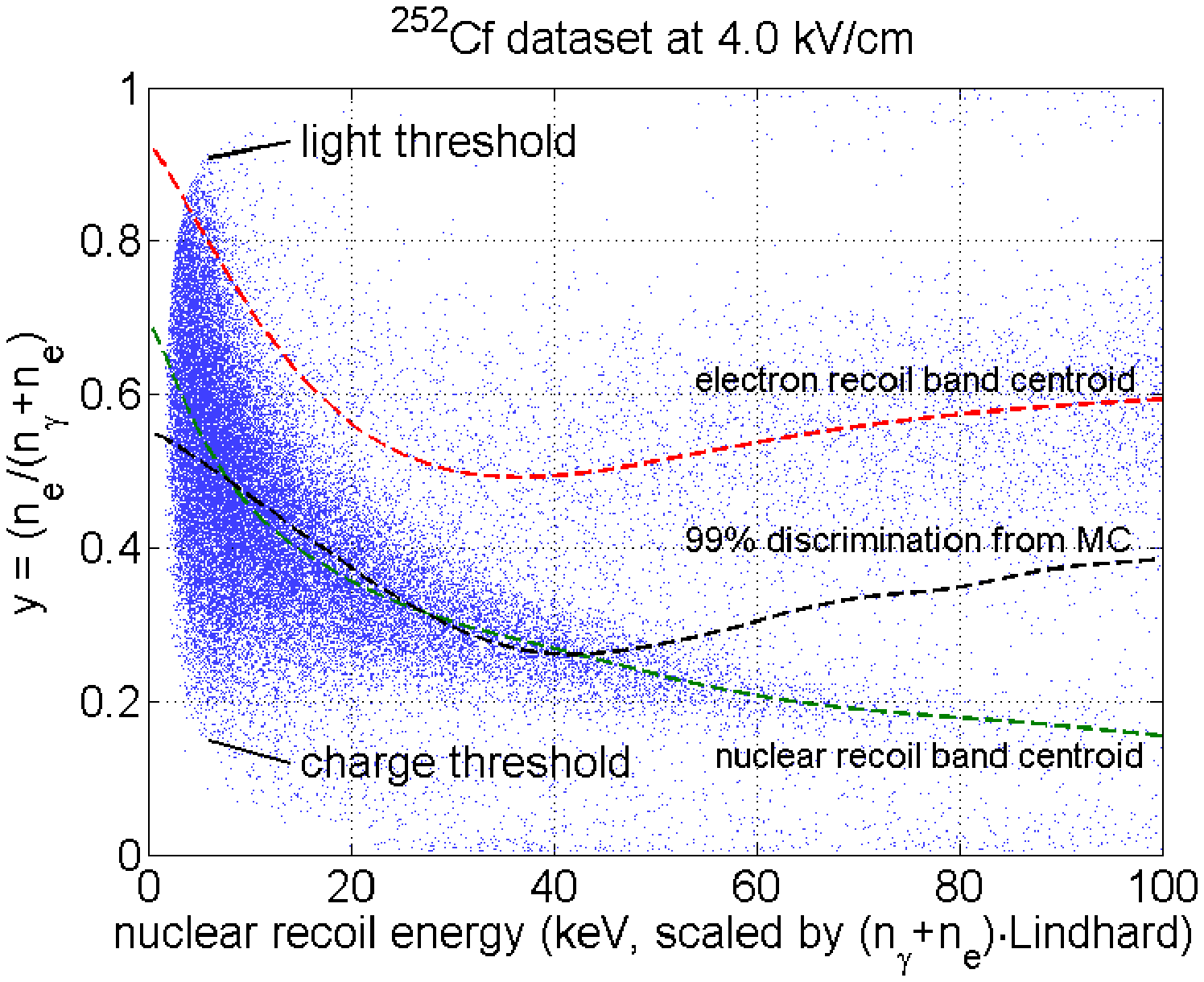}
	\includegraphics[width =3.25in]{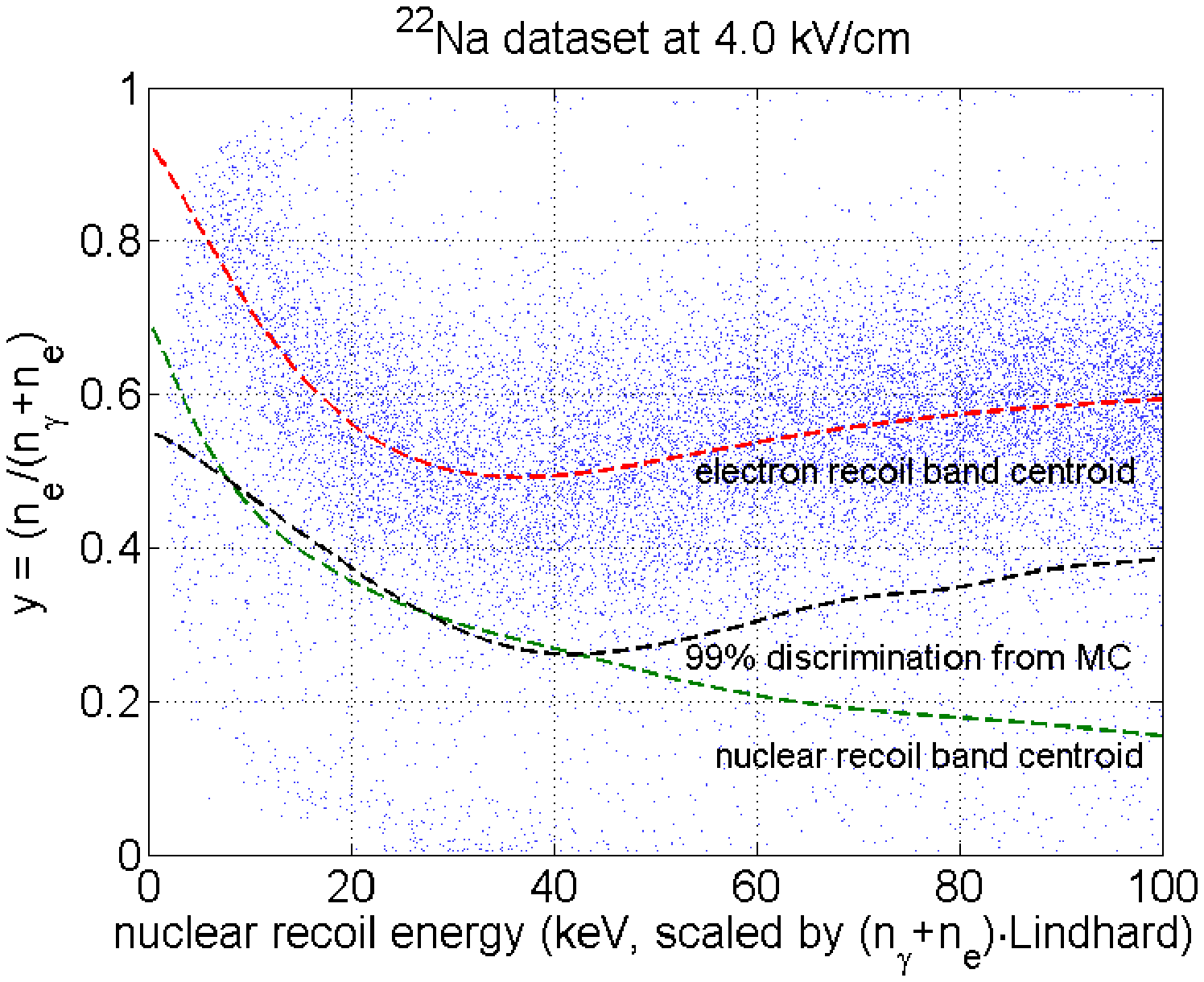}
\end{center}
\caption{\label{fig:LowEDiscrim}Discrimination at low energies illustrated by the response of the detector to \emph{below} compton-scatters of gammas from $^{22}$Na, and \emph{above} nuclear recoils and gammas from a $^{252}$Cf neutron source. The curves show the centroids of the bands (ignoring tails), and the 99\% rejection contour deduced from MonteCarlo.}

\end{figure}

Perhaps the most important characteristic of this class of detectors is their ability to distinguish electron and nuclear recoils based on the greater recombination for nuclear recoils.   We form the discrimination parameter $y=n_e/(n_e+n_\gamma)$, the fraction of the total signal seen in the charge channel, and plot this versus electron and nuclear recoil energies in Figs. \ref{fig:Bands_w_40_keV} and \ref{fig:LowEDiscrim}.  We also find the centroids of the electron and nuclear recoil bands versus energy, given by Compton scatters and elastic scatters from neutrons, respectively.  For multiple-recoil events, such as photo-absorption of gammas and inelastic nuclear recoils, $y$ is simply the weighted average of the contributing recoils.  This provides a useful check on the calibration, as we can compare the positions of the prompt 40~keV inelastic nuclear recoil and 122~keV photo-absorption peaks with those predicted from the simple recoil bands, as shown in Fig.~\ref{fig:Bands_w_40_keV}.

The discrimination can be measured directly from the data in Fig.  \ref{fig:LowEDiscrim}, and is $\gtrsim$98\% above 20 keVr, falling to 95\% at 10 keVr.   However most of the limitation to this appears to be a diffuse `tail' of electron recoil events with poor charge collection that we believe are due to charge loss at the edges of the active region.   This will be eliminated by x-y resolving detectors such as XENON10.  We deduce the limiting discrimination using a Monte Carlo simulation including photon collection statistics, the measured resolutions in S1 and S2, and an energy-dependent recombination fluctuation term tuned to match the width of the main portion of the electron recoil bands.  The resulting 99\% discrimination curve is shown.  For most of the energy range shown and with reasonable assumptions about the energy-dependence of the S1 and S2 instrumental resolutions, recombination fluctuations are the dominant limitation to discrimination.    Finally, in this data, the first such measurement below 20 keVr, we see a striking increase in $y$ for electron recoils with E~$\lesssim$~35~keVr (and to a lesser extent nuclear recoils with E~$\lesssim$~20~keVr).  The origin of this is not yet understood, but its importance in enabling discrimination to below 10 keVr cannot be overstated.   The WIMP sensitivity of Xe drops steeply with energy due to the nuclear form factor \cite{Lewin:1995factorreview}; thus these results showing promise for low energy discrimination demonstrate the power of this new class of detectors.

\bibliography{Sorma}
%
%
%
%
%

\end{document}